# The origin of correlation fringe in Franson-type experiments

Byoung S. Ham

Center for Photon Information Processing, School of Electrical Engineering and Computer Science,
Gwangju Institute of Science and Technology, 123 Chumdangwagi-ro, Buk-gu, Gwangju 61005, S. Korea
(Submitted on May 29, 2020)

Nonlocal correlation is the key concept in quantum information processing, where quantum entanglement provides such a nonclassical property. Since the first proposal of noninterfering interferometer-based two-photon intensity correlation by Franson (Phys. Rev. Lett. **62**, 2205 (1989)), the particle nature of photons has been intensively studied for nonlocal correlation using Mach-Zehnder interferometers (MZIs). Here, the role of MZIs is investigated with respect to the origin of nonlocal correlation in Franson-type experiments, where the wave nature of photons plays a critical role. Under the coincidence-provided quantum superposition between independent MZIs, we prove that nonlocal correlation can be created from non-entangled photons through the MZIs.

*Introduction*

Nonlocal correlation [1-21], the quintessence of quantum mechanics, has made quantum computing [4], quantum communications [5-7], and quantum sensing [8,9] unique compared with their classical counterparts. Since violation of Bell's inequality had been reported from both phase [10] and polarization bases [11], Franson proposed a different type of nonlocal correlation using energy and time conjugate relations satisfying Heisenberg uncertainty principle [12]. For Franson-type experiments, a photon pair entering two independent Mach-Zehnder interferometers (MZIs) is manipulated to be indistinguishable for two-photon coincidence detection, violating Bell's inequality in a very delicate manner presenting as an interference fringe [11-20]. In the Franson-type nonlocal correlation, however, both photons passing through individual MZIs never interfere with each other until measured independently by separate photon detectors. Moreover, each MZI is designed to be prevented from self-interference for each single photon by an asymmetric delay-line setup exceeding coherence length of the photons. Thus, the observed intensity fringe violating Bell's inequality in the Franson-type experiments is mysterious and differentiated from non-fringe correlations observed in Hong-Ou-Mandel (HOM)-type experiments with two input photons interfering on a BS [21] and Hanbury Brown and Twiss (HBT) experiments based on a beam splitter (BS) with one input [22].

The common ground of nonlocal correlation and quantum entanglement is the quantum superposition-caused indistinguishability between two photons for coincidence detection satisfying uncertainty principle of quantum mechanics [23]. Recently, the origin of nonclassicality observed in HOM-type experiments has been investigated and found in a phase-locked, photon-pair relationship [24]. Unlike common understanding of the particle nature of photons for coincidence detection where the phase term must be neglected, the coherence analysis in ref. 24 makes it clear and important, where the same physics can also be applied for the observed trapped ion entanglement [25]. In other words, the anticorrelation in HOM-type experiments is now understood as the wave nature of photons under quantum superposition provided by coincidence detection. Thus, the Franson-type correlation fringe might be explained in such a way. The HBT-type experiments, however, have nothing to do with interference because there is no chance for incoming photons to be interfered. Thus, the nonclassical feature observed in HBT-type experiments is simply due to anti-bunched photons [26,27].

Here, we analyze the Franson-type experiments to investigate the origin of the correlation fringe demonstrated in a noninterfering interferometry. For this, usual particle nature of coincidence detection is analyzed, in which quantum superposition between coincidently detected photons becomes the bedrock of the interference fringe. Because the two MZIs never interfere with each other, the observed fringe seems to be some weird involvement of the wave nature. Coincidence detection is a necessary condition for fringe formation in terms of fulfilling two-photon correlation by definition within the coherence time. More importantly, however, the coincidence detection induces quantum superposition (indistinguishability) between two photons passing through different route combinations in the two independent MZIs. Thus, the origin of correlation fringe is rooted in the noninterfering interferometers via coincidence detection-provided path superposition. In other words, the pair of photons entering the two independent MZIs becomes interfered with each other in some way, resulting in the fringe as a function of path length difference. The role of noninterfering interferometers for photon pair correlation is therefore key to understanding the nonlocal correlation. Franson-type experiments should belong to such a category of coherence optics under the particle nature of coincidence detection, satisfying wave-particle duality [28].

*Analysis*

Figure 1 shows a schematic of a typical Franson-type experimental setup, where the noninterfering interferometers measure the intensity correlation $g_{AB}^{(2)}$ between two output photons, $E_A$ and $E_B$. In Fig. 1, the input photons of $E_1$ at center frequency of $f_1$ and $E_2$ at center frequency of $f_2$ are supposed to be antibunched satisfying sub-Poisson photon statistics. A typical setup for the Franson-type experiments is to use two MZIs with asymmetric stucture results in no (self) interference to satisfy the particle nature of photons in terms of separability (see Figs. 1(a) and (b)). Suppose that the light source S generates photon pairs of $E_1$ and $E_2$ by, e.g., spontaneous parametric down conversion (SPDC) processes, in which each photon pair is strongly correlated with each other by the sum frequency lock at $f_0$ according to the energy conservation law [29]: $f_0 = f_1 + f_2$ (see Figs. 1(c) and (d)). To satisfy lack of interference in each MZI in Fig. 1, the time delay $\Delta t_j\ (= \frac{1}{c}(L_j - S_j); j = 1,2)$ between the short and long paths must be longer than the photon coherence time $\tau_C\ (= \frac{1}{\Delta f})$, where $\Delta f$ is the spectral bandwidth of the light source S. According to the Heisenberg uncertainty principle, the condition of $\frac{\Delta f \Delta L_j}{c} \geq 1$ must be satisfied. For a symmetric MZI ($\Delta t_j \ll \tau_C$), the MZI output probability of $E_A$ ($E_B$) is independent of the phase or frequency fluctuations of the input photons $E_1$ ($E_2$) [30]. Thus, the spectral bandwidth of S becomes a free parameter to each MZI in terms of interference (wave nature) except for functions of photon separability (particle nature). Figures 1(c) and (d) is to show the difference between degenerate and nondegenerate SPDC processes (analyzed in Fig. 2).

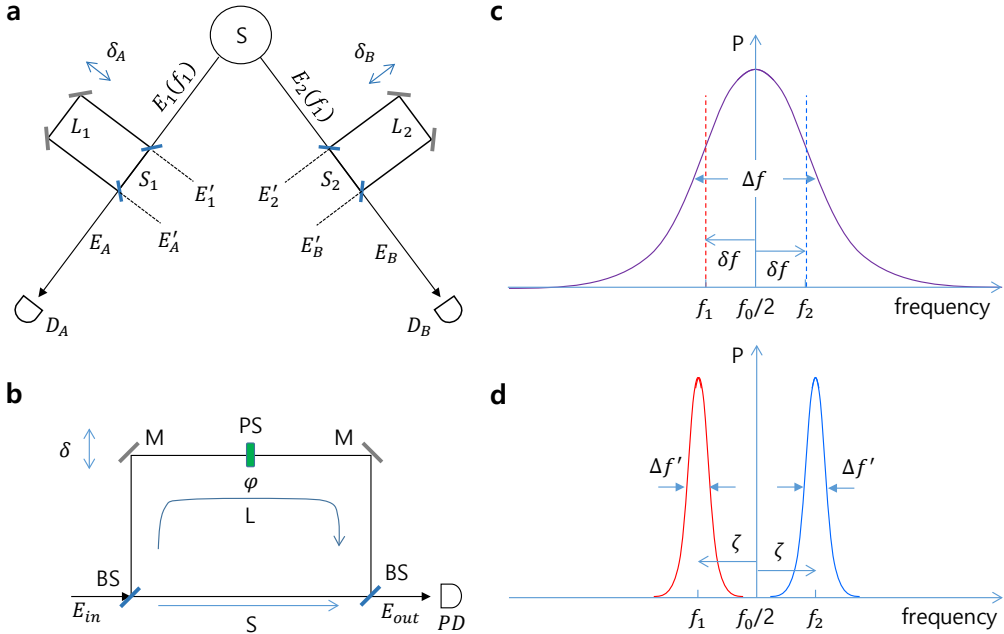

Fig. 1. A schematic of Franson-type intensity correlation. (a) The asymmetric MZI-based Franson-type setup. (b) Details of each MZI where two paths (L and S) never interfere, resulting in photon separation in the output modes $E_A$ and $E_B$. (c) A degenerate scheme of SPDC. (d) A nondegenerate scheme of SPDC. S: Entangled photon source. The δ represents path length control with long path L, where the phase shifter can replace the δ. BS: beam splitter, M: mirror, PD: photon detector, PS: phase shifter (replaced by δ), $\Delta f / \Delta f'$: photon bandwidth of S,

In the Franson-type setup of Fig. 1, the asymmetric path configuration in each MZI satisfies the complete particle nature of photons with $\delta_{A,B}(\equiv L_{1,2} - S_{1,2}) \gg l_C$. This condition leads to an orthogonal basis relationship between short and long paths, $|S\rangle$ and $|L\rangle$, resulting in no (self) interference with $\langle S|L\rangle$=0. In general, each photon can be described as a traveling wave, whose wave vector and initial phase are denoted by $\varphi_{S1}$ for $E_1$ and $\varphi_{S2}$ for $E_2$, respectively. Keeping this in mind, the following results are obtained for the first-order correlation $g_{A,B}^{(1)}$ in each MZI in Fig. 1, where $E_1$ and $E_2$ do not need to be entangled for this analysis:

(i) 
$$E_A = \frac{1}{\sqrt{2}}E_0(e^{i\varphi_{S1}}|S\rangle + e^{i\varphi_{L1}}|L\rangle)e^{-i\omega_1 t}$$
$$= \frac{1}{\sqrt{2}}E_0 e^{i\varphi_{S1}}(|S\rangle + e^{i(\varphi_{L1}-\varphi_{S1})}|L\rangle)e^{-i\omega_1 t}$$
$$= \frac{1}{\sqrt{2}}E_0 e^{i\varphi_{S1}}(|S\rangle + e^{i\varphi_A}|L\rangle)e^{-i\omega_1 t}. \quad (1)$$

$$I_A = E_A E_A^* = \frac{1}{2}I_0[\langle S|S\rangle + \langle L|L\rangle + \langle S|L\rangle(e^{i\varphi_A} + e^{-i\varphi_A})]$$
$$= I_0[1 + \langle S|L\rangle\cos(\varphi_A)]. \quad (2)$$

(ii)
$$E_B = \frac{1}{\sqrt{2}}E_0(e^{i\varphi_{S2}}|S\rangle + e^{i\varphi_{L2}}|L\rangle)e^{-i\omega_2 t}$$
$$= \frac{1}{\sqrt{2}}E_0 e^{i\varphi_{S2}}(|S\rangle + e^{i(\varphi_{L2}-\varphi_{S2})}|L\rangle)e^{-i\omega_2 t}$$
$$= \frac{1}{\sqrt{2}}E_0 e^{i\varphi_{S2}}(|S\rangle + e^{i\varphi_B}|L\rangle)e^{-i\omega_2 t}. \quad (3)$$

$$I_B = E_B E_B^* = \frac{1}{2}I_0[\langle S|S\rangle + \langle L|L\rangle + \langle S|L\rangle(e^{i\varphi_B} + e^{-i\varphi_B})]$$
$$= I_0[1 + \langle S|L\rangle\cos(\varphi_B)], \quad (4)$$

where $\varphi_A = \varphi_{L1} - \varphi_{S1}$, $\varphi_B = \varphi_{L2} - \varphi_{S2}$, $\varphi_{Lj} = \frac{2\pi}{c}f_j L_j$, and $\varphi_{Sj} = \frac{2\pi}{c}f_j S_j$. The subscripts 1 and 2 indicates different MZIs with different photons. As a result, the amplitude correlation in Eqs. (2) and (4) for each MZI output photon has no interference fringe because of $\langle S|L\rangle = 0$. This result in each interferometer is consistent with photon characteristics of $E_1$ and $E_2$, whether they are entangled (quantum) or not (classical).

Now, two output photons $E_A$ and $E_B$ are coincidently detected, and their intensity correlation $g_{AB}^{(2)}(0)$ is obtained from Eqs. (1) and (3):

$$g_{AB}^{(2)}(0) = \frac{\langle E_A E_A^* E_B E_B^*\rangle}{\langle I_A\rangle\langle I_B\rangle}$$

$$= \frac{1}{4}\langle(|S\rangle + e^{i\varphi_A}|L\rangle)(\langle S| + e^{i\varphi_B}\langle L|)(\langle S| + e^{-i\varphi_A}\langle L|)(|S\rangle + e^{-i\varphi_B}|L\rangle)\rangle$$

$$= \frac{1}{4}\langle(\langle S|S\rangle + e^{i(\varphi_A+\varphi_B)}\langle L|L\rangle)(\langle S|S\rangle + e^{-i(\varphi_A+\varphi_B)}\langle L|L\rangle)\rangle$$

$$= \frac{1}{2}\langle 1 + \cos\left[\frac{\pi}{c}f_0(\Delta L_1 + \Delta L_2) \pm \frac{2\pi}{c}\delta f(\Delta L_1 - \Delta L_2)\right]\rangle, \quad (5)$$

where $\varphi_A = \frac{2\pi}{c}f_1\Delta L_1$, $\varphi_B = \frac{2\pi}{c}f_2\Delta L_2$, $\Delta L_1 = L_1 - S_1$, and $\Delta L_2 = L_2 - S_2$. In Eq. (5), the condition of coincidence detection is $l_c \left(= \frac{c}{\Delta f}\right) > (\Delta L_1 - \Delta L_2)$. Because the short paths ($S_1$ and $S_2$) are fixed, this condition becomes the function of long-path difference. For degenerate SPDC processes (see Fig. 1(c)), the frequencies of signal ($f_1$) and idler ($f_2$) photons are symmetrically detuned across $f_0/2$ within the bandwidth $\Delta f$, and $\delta f$ is a symmetric detuning of each photon pair from $f_0/2$ by the energy conservation law of $f_0 = f_1 + f_2$, where $f_1 = \frac{f_0}{2} \pm \delta f$, $f_2 = \frac{f_0}{2} \mp \delta f$, and $\delta f \leq \Delta f$. For coincidence detection ($\tau \sim 0$), the second term in the parentheses of Eq. (5), $\frac{\pi}{c}\delta f(\Delta L_1 - \Delta L_2)$, is nearly zero due to $\Delta L_1 \sim \Delta L_2$ ($\sim \Delta L$) within $\tau < \tau_c$ by $\frac{\Delta f}{c}(\Delta L_1 - \Delta L_2)_{max} < 1$ (see also Figs. 2(a) and (b)). This frequency lock with symmetric detuning is the most important condition for the formation of the fringe. Even for the nondegenerate SPDC case (see Fig. 1(d)), Eq. (5) still works with an additional but negligible term of $\frac{2\pi}{c}\zeta(\Delta L_1 - \Delta L_2)$, where $f_1 = \frac{f_0}{2} + \zeta \pm \delta f'$ and $f_2 = \frac{f_0}{2} - \zeta \mp \delta f'$ [16]. The only issue is a reduced range in path length difference by $\zeta$: $\delta f = \zeta + \delta f'$. Thus, the intensity correlation becomes only a function of $\cos\left(\frac{2\pi}{c}f_0\Delta L\right)$, where $\langle\cos\left(\frac{2\pi}{c}f_0\Delta L\right)\rangle \sim \cos\left(\frac{2\pi}{c}f_0\Delta L\right)$ if $\Delta L < l_c$. Here, the linewidth ($\delta f_0$) of the pump laser ($f_0$) is generally taken to be $\delta f_0 \ll \Delta f$. This status of $\delta f_0 \ll \Delta f$ is the second most important condition for good fringe formation exceeding classical bounds. Under these conditions, Eq. (5) can be rewritten as (see Fig. 2):

$$g_{AB}^{(2)}(0) = \frac{1}{2}\left[1 + \cos\left(\frac{2\pi}{c}f_0 \Delta L\right)\right]. \tag{6}$$

Although each MZI is prohibited from self-interference, the second-order correlation $g_{AB}^{(2)}(0)$ in Eq. (6) creates an interference fringe as a function of $\Delta L$ within the coherence length. For a fixed $\Delta L_1$ and each short path length $S_1$ and $S_2$, $2\Delta L = (\Delta L_1 + \Delta L_2) = 2\Delta L_1 + (L_2 - L_1)$. Thus, the physical origin of Eq. (6) is in, first, the sum frequency lock at $f_0$; second, the oppositely detuned frequencies between two input photons whose detuning bandwidth is the source bandwidth of $\Delta f$; and third, a narrower pump laser linewidth $\delta f_0$, satisfying $\delta f_0/\Delta f \ll 1$. For an extreme case of $\Delta L \gg l_c$, $g_{AB}^{(2)}(\tau)$ converges to 0.5 as shown in Eq. (6) as well as in Fig. 2(b), which indicates the classical lower bound for completely independent photons [24,31].

Figure 2 shows numerical calculations for Eq. (5) using experimental parameters observed in Franson-type experiments. For simplicity, however, we set the wavelength of the pump photon at 1μm. In Fig. 2, the pump laser linewidth effect is simply neglected due to $\frac{\delta f_0}{\Delta f} \ll 1$. The red curve in Fig. 2(a) is a typical Franson-type fringe observed for $\Delta L \ll l_C$. The correlation fringe exists until $\Delta L \sim l_C$ as shown in Fig. 2(b). In Fig. 2(a), nonsymmetric cases (green, blue, and dotted curves) are also compared with the symmetric case (red curve), where the $\Delta L_1 - \Delta L_2$ term is simply replaced by $\Delta L_1 + \Delta L_2$ for the worst case with $|f_1 - f_2| \neq 2\delta f$. If the photon bandwidth of S can be engineered to be less than 1GHz or so, not only the fringe, but also Bell's inequality violation can still be achieved as shown in the green curve in Fig. 2(a). The fringe is a result of $\Delta L$ to $l_C$, where the coherence washout effect can be negligibly small even for the nonsymmetric case. Here, the nonsymmetric case can be easily manipulated using an etalon from sunlight or other independent light sources [32,33]. Thus, the existence of the green curve in Fig. 1 newly defines the nonlocal correlation observed in Franson-type experiments. In Figs. 2(c) and 2(d), the coherence washout appears at $\Delta f = 2GHz$ (see the dashed lines for $\delta f$ sum). Thus, the Franson-type fringe can also be obtained from an independent light source regardless of entanglement. In other words, the independent photons become entangled through MZIs under coincidence-induced quantum superposition. This is the newly discovered discrete function of MZI in Franson-type experiments.

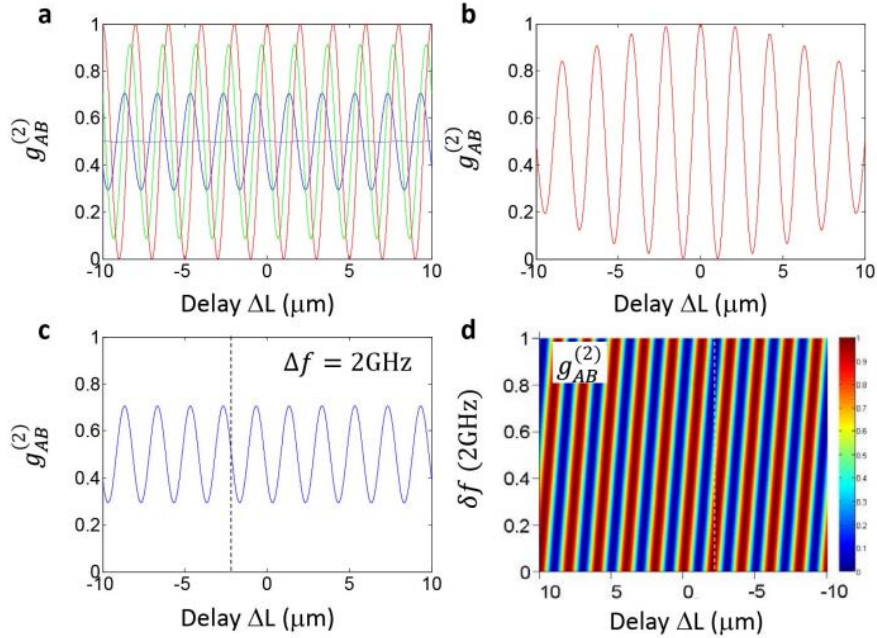

Fig. 2. Numerical calculations of Eq. (5). $g_{AB}^{(2)}(\tau)$ for symmetric detuning (a) $\Delta f = 3GHz$, and (b) $\Delta f = 20THz$. (c) and (d) $g_{AB}^{(2)}(0)$ for nonsymmetric detuning for the blue curve in (a) with $\Delta f = 2GHz$. $f_0 = 300\ THz\ (\lambda_0 = 1\mu m)$; $\Delta L_1 = 0.1m$; $\Delta L = \Delta L_1 - \Delta L_2$, where (c) is the sum of (d) along $\delta f$. Fig. In (a), the following are for nonsymmetric detuning: Green, $\Delta f = 1GHz$; Blue, $\Delta f = 2GHz$; Dotted(center line), $\Delta f = 3GHz$. In (c) and (d), the dashed lines are for $\Delta f$-induced decoherence.

*Discussion*

The origin of the correlation fringe in Franson-type experiments has been analyzed with the wave nature of photons whose fringe is due to the sum frequency locking of two input photons with symmetric frequency detuning. The role of coincidence detection is to provide quantum superposition between two possible path combinations in independent interferometers, resulting in indistinguishability between coincidently detected photon pairs. Thus, the two-photon intensity correlation is an amplitude correlation between coincidence detection-caused coupled MZIs. Without violating the Heisenberg uncertainty principle between time and energy variations, the sum frequency lock of the two input photons results in the accurate measurement of coherence for a fringe due to the symmetric detuning-caused noise cancellation. Although this symmetric (opposite) detuning with the sum frequency lock plays a key role in the fringe formation in Eq. (5), individual (classical) photons can also result in the fringe via photon engineering with either an unlocked sum-frequency within the narrow bandwidth (GHz) or a locked sum frequency with wide bandwidth of ~30THz (100nm) regardless of degeneracy in photon pair generation [16]. Thus, Franson-type fringe visibility can also be effective on independent (classical) input photons. In that sense, independent MZI interferometers should work for entanglement generation under the coincidence detection-provided quantum superposition environment.

*Conclusion*

In summary, the interference fringe observed in Franson-type experiments was analyzed and discussed with the wave nature of photons under the coincidence detection-provided quantum superposition environment. The superposition-caused coherence between noninterfering photons was analyzed with sum frequency locked photon pairs. The sum frequency lock provides simultaneous accuracies in both frequency and time without violating the Heisenberg uncertainty principle. For fringe analysis, the frequency bandwidth-caused coherence washout was also analytically and numerically demonstrated for both coherence washout and non-washout depending on photon bandwidth. Symmetric detuning of two input photons with sum-frequency lock can also be obtained from four-wave-mixing-caused bi-photon generations [16,34,35]. The classical lower bound of $g_{AB}^{(2)}(0) = 0.5$ was also demonstrated when coherence between two photons was lost by delaying one photon relative to the other. Although the Franson-type setup has been used to prove nonclassical features of input photons with Bell's inequality violation, the nonclassical visibility fringe can also be applied to classical input photons. Thus, the interferometer should act as a quantum device to generate nonlocal correlation even from classical counterparts.

*Acknowledgement:*

This work was supported by a GIST research institute (GRI) grant funded by GIST in 2020.